\documentclass[11pt]{article}
\usepackage[letterpaper,hmargin=1in,vmargin=1.25in]{geometry}
\usepackage[english]{babel}
\usepackage[utf8]{inputenc}
\usepackage[T1]{fontenc}
\usepackage{amsmath}
\usepackage{amsfonts}
\usepackage{amssymb}
\usepackage{framed}
\usepackage{dsfont} 
\usepackage[standard]{ntheorem}
\usepackage[dvips]{graphics}
\usepackage{epsfig}
\usepackage{epsfig,psfrag}
\usepackage{subfigure}
\usepackage{color}
\usepackage{calc}
\usepackage{listings}
\usepackage{url}
\usepackage{alltt}
\usepackage{tabls}
\usepackage{textcomp}
\usepackage{slashbox}
\usepackage{times}
\usepackage{tabularx}
\usepackage{sublabel}

\title{A topological chaos framework for watermarking}
\author{Jacques M. Bahi, Christophe Guyeux.\\\\
\emph{Laboratoire d'Informatique de Franche-Comte,}\\
\emph{Universite de Franche-Comte, France.}\\
\href{mailto:bahi@univ-fcomte.fr}{bahi@univ-fcomte.fr}, ~\href{mailto:guyeux@iut-bm.univ-fcomte.fr}{guyeux@iut-bm.univ-fcomte.fr}}

\begin{document}

\title{A watermarking algorithm satisfying topological chaos properties}
\author{Jacques M. Bahi, Christophe Guyeux. \\
\\
\emph{Laboratoire d'Informatique de Franche-Comté,}\\
\emph{Université de Franche-Comté, France}\\
\href{mailto:bahi@univ-fcomte.fr}{jacques.bahi@univ-fcomte.fr}, ~\href{mailto:guyeux@iut-bm.univ-fcomte.fr%
}{guyeux@iut-bm.univ-fcomte.fr}}
\maketitle

\begin{abstract}
A new watermarking algorithm is given, it is based on the so-called chaotic
iterations and on the choice of some coefficients which are deduced from the
description of the carrier medium. After defining these coefficients,
chaotic discrete iterations are used to encrypt the watermark and to embed
it in the carrier medium. This procedure generates a topological chaos and
ensures that the required properties of a watermarking algorithm are
satisfied.
\end{abstract}

\setcounter{secnumdepth}{3}

\emph{Key-words: Watermarking, Encryption, Chaotic iterations, Topological
chaos, Information hiding}

\section{Introduction}

Information hiding has recently become a major information security
technology, especially with the increasing importance and widespread
distribution of digital media through internet. Its aim is to embed a piece
of information into digital documents, like pictures or movies, for a large
panel of reasons, such as copyright protection, control utilization, data
description, content authentication, and data integrity. Digital
watermarking is one the techniques used in information hiding. It must offer
a lot of desirable characteristics, including security, imperceptibility and
robustness. To do so, many different watermarking schemes have been proposed
in recent years, which can be classified into two categories: spatial
domain, and frequency domain watermarking. In spatial domain watermarking, a
large number of bits can be embedded without incurring noticeable visual
artifacts; whereas, frequency domain watermarking has been shown to be quite
robust against JPEG compression, filtering, noise pollution and so on. More
recently, chaotic methods have been proposed to encrypt the watermark before
embedding it in carrier image, for security reasons. This methods are
usually based on three fundamental chaotic maps: Chebychev, logistic and
Arnold's cat maps. 
%
%
%
%

\medskip 

I In this paper, a new watermarking algorithm is given, it is based on the
so-called chaotic iterations and on the choice of some coefficients which
are deduced from the description of the carrier medium. After defining these
coefficients, chaotic discrete iterations are used to encrypt the watermark
and to embed it in the carrier medium. 

The new algorithm consists in two stages, an encryption stage which
encompasses many encryption algorithms and makes them more secure and an
embedding stage, also based on chaotic iterations. An authentication of the
relevant information carried by the support, can be done during each of this
two stages.

The algorithm generates a topological chaos in the sense of Devaney and
ensures by this way that the required properties of a watermarking algorithm
are satisfied.

After introducing the new algorithm and explaining its theoretical
foundations, a case study allows us to evaluated it.

\medskip

The rest of this paper is organized as follows: first, some basic
definitions concerning chaotic iterations are recalled. Then, the new
chaos-based watermarking algorithm is introduced. The next section is
devoted to the related works and to the contribution of our results in light
of existing ones. Section $5$ is devoted to the evaluation of our algorithm.
A case study is presented and some classical attacks are executed, the
results are presented and commented. The paper ends by a conclusion section
where the contributions of the paper are summarized and the planned future
work are given.

\section{Chaotic iterations: basic recalls}

\label{chaotic iterations} In the sequel $[| 1;N|]$ means $%
\{1,2,\hdots,N\}$, $s^{n}$ denotes the $n^{th}$ term of a sequence $%
s=(s^{1},s^{2},...)$, $V_{i}$ denotes the $i^{th}$ component of a vector $%
V=(V_{1},V_{2},...)$, and $f^{k}$ denotes the $k^{th}$ composition of a
function $f,$ $f^{k}=\underset{k\text{ }times}{\underbrace{f\circ f\circ
...\circ f}}$.

Let us consider a \emph{system} of a finite number $\mathsf{N}$ of \emph{%
cells} so that each cell has a boolean \emph{state}. Then a sequence of
length $\mathsf{N}$ of boolean states of the cells corresponds to a
particular \emph{state of the system}.

\begin{definition}
A \emph{chaotic strategy} corresponds to a sequence of $[|1;\mathsf{N%
}|]$. The set of all chaotic strategies is denoted by $\mathcal{S}.$
\end{definition}

\begin{definition}
\label{Def:Chaotic iterations}$\mathds{B}$ denoting $\{0,1\}$, let $f:%
\mathds{B}^{\mathsf{N}}\longrightarrow \mathds{B}^{\mathsf{N}}$ be an
iteration function and $S\in \mathcal{S}$ be a chaotic strategy. The so
called \emph{chaotic iterations} are defined by (see \cite{Robert1986})%
\begin{equation}
\left\{ 
\begin{array}{l}
x^{0}\in \mathds{B}^{\mathsf{N}} \\ 
\forall n\in \mathds{N}^{\ast },\forall i\in [|1;\mathsf{N}|]%
,x_{i}^{n}=\left\{ 
\begin{array}{ll}
x_{i}^{n-1} & \text{ if }S^{n}\neq i \\ 
f(x^{n})_{S^{n}} & \text{ if }S^{n}=i.%
\end{array}%
\right. 
\end{array}%
\right.   
\end{equation}%
In other words, at the $n^{th}$ iteration, only the $S^{n}-$th cell is
\textquotedblleft iterated\textquotedblright . Note that in a more general
formulation, $f(x^{n})_{S^{n}}$ can be replaced by $f(x^{k})_{S^{n}}$, where 
$k\leqslant n$, modelizing for example delays between cells (see \emph{e.g.} 
\cite{Bahi2000}).
\end{definition}

Chaotic iterations generate a set of vectors (boolean vector in this paper),
they are defined by an initial state $x^{0},$ an iteration function $f$ and
a chaotic strategy $S.$

\section{A new chaos-based watermarking algorithm}

\subsection{Most and least significant coefficients}

Let us first introduce the definition of most and least significant
coefficient of an image.

\begin{definition}
For a given image, most significant coefficients (in short MSCs), are
coefficients that allow the description of the relevant part of the medium, 
\emph{i.e.} its most rich part (in terms of embedding informations), through
a sequence of bits.
\end{definition}

For example, in a spatial description of a grayscale image, a definition of
MSCs can be the sequence constituted by the three first bits of each pixel.
In a frequency discrete cosine domain description, each $8\times 8$ block of
the carrier image is mapped to a list of 64 coefficients; the energy of the
image is contained in the firsts of them, so the first fourth coefficients
of all the blocks can be a good sequence of MSCs, after binary conversion.

\begin{definition}
By least significant coefficients (LSCs), we mean a translation of some
unsignificant parts of a medium in a sequence of bits (unsignificant can be
understand as: ``which can be altered without sensitive damages'').
\end{definition}

This LSCs can be, for example, the last three bits of the gray level of each
pixel, in case of spatial domain watermarking.

\begin{center}
\begin{tabular}{cc}
\includegraphics[scale=0.5]{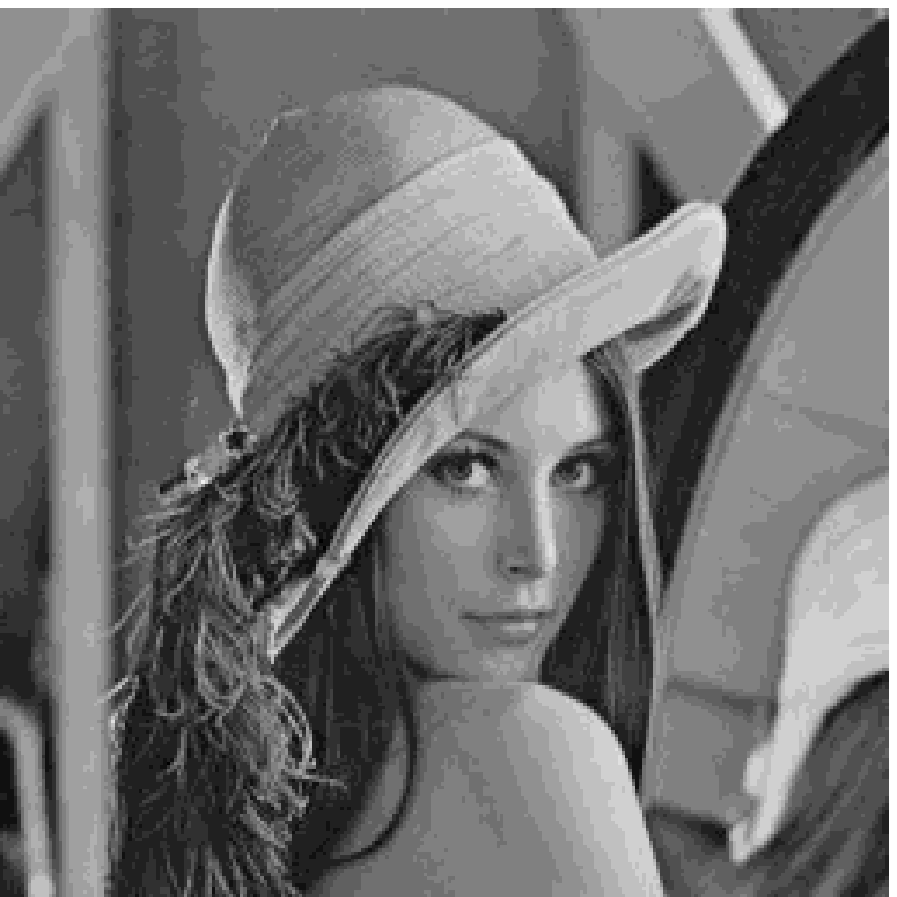} & %
\includegraphics[scale=0.5]{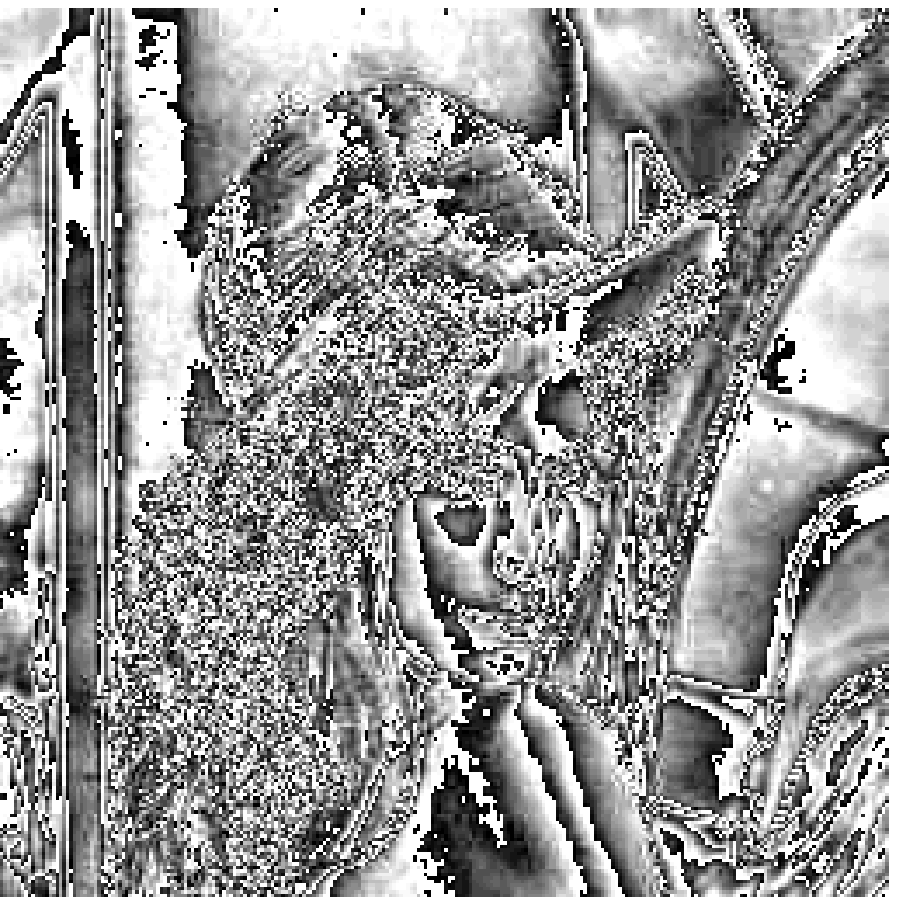} \\ 
\emph{Fig 1a.} MSCs of Lena & \emph{Fig 1b.} LSCs of Lena, \\ 
& pixel value increased by a factor 10%
\end{tabular}
\end{center}

Discrete cosine, Fourier and wavelet transform can be used to generate LSCs
and MSCs, in case of frequency domain watermarking, but other choices are
possible.

\medskip

LSCs are used during the stage of embedding: some of the least significant
coefficients of the carrier image will be chaotically chosen, and replaced
by the bits of the (possibly encrypted) watermark. With a tall number of
LSCs, the watermark can be inserted more than one time, and thus the
embedding will be more secure and robust, but more detectable.

MSCs are only useful in case of authentication: thus, encryption and
embedding will depend on them. As a consequence, a coefficient should not be
defined, at the same time, as a MSC and a LSC: the latter can be altered,
while the former are needed to extract the watermark (in case of
authentication).

\subsection{Stages of the algorithm}

Our watermarking scheme consists in two stages: encryption of the watermark
and its embedding.

\subsubsection{Watermark encryption}

For security reasons, the watermark can be encrypted before its embedding
into the image. A common way to achieve this stage is to use the bitwise
exclusive or (XOR), for example between the watermark and a logistic map%
\footnote{%
The Logistic map is defined by 
\begin{equation*}
\begin{array}{ll}
U^{0}\in ]0,1[,~\mu \in \lbrack 3.57;4], &  \\ 
U^{n+1}=\mu ~U^{n}~(1-U^{n}). & 
\end{array}%
\end{equation*}%
To obtain a bit sequence $X$, the following proceeding can be applied: if $%
U^{k}<0.5$, then $X^{k}=0$, else $X^{k}=1$.}. In this document, we will
introduce a new encryption scheme, based on chaotic iterations. Its chaotic
strategy will be high sensitive to the MSCs, in case of an authenticated
watermark. For more precision see\ref{watermark encryption} in section \ref%
{CaseStudy} below.

\subsubsection{Watermark embedding}

Some LSCs will be substituted by the bits of the possibly encrypted
watermark. To choose the sequence of LSCs to be replaced, a chaotic sequence 
$\left( U^{k}\right) _{k}$ of integers, lower than the number $N$ of LSCs,
is generated from the chaotic strategy used in the encryption stage. Thus,
the $U^{k}-th $ least significant coefficient of the carrier image is
substituted by the $k^{th}$ bit of the possibly encrypted watermark. In case
of authentication, such a procedure conducts to a choice of the LSCs highly
dependant to the MSCs. See\ref{Watermark embedding} in section \ref%
{CaseStudy} for more details.

\subsection{Extraction}

The chaotic strategy can be regenerated, even in the case of an
authenticated watermarking: the MSCs have not been changed during the stage
of embedding watermark. Thus, the few altered LSCs can be found, and the
encrypted watermark can be rebuilt, and decrypted.

If the watermarked image is attacked, then the MSCs will change.
Consequently, \emph{in case of authentication }and\emph{\ }due to the high
sensitivity of the embedding sequence, the LSCs designed to receive the
watermark will be completely different. Hence, the result of the decrypting
stage of the extracted bits will have no similarity with the original
watermark.

\subsection{The general chaos-based watermarking algorithm}

The different stages of our method can be described in a more general
framework. First of all, the representation domain of the carrier image has
to be decided (spatial, DCT, DWT, \emph{etc.}) As explained above, this
choice will affect the following MSCs and LSCs (most and least significant
coefficients). Then, the question of the encryption of the watermark must be
asked, and if an encryption is needed, then a cipher method will be chosen.
In this paper, the watermark has been encrypted with chaotic iterations, but
other methods can be chosen, such as making the bitwise exclusive or with
some other boolean maps.

\medskip

For the next stage, we need to know whether the embedding has to be
authenticated. If so, the significant information contained in the carrier
image must be summarized in a selection of relevant MSCs: there are lot of
possibilities, depending on the representation domain. Then, in case of
authentication, the MSCs must be associated, on the one hand, to the
encryption of the watermark, and on the other hand, to the embedding of the
possibly encrypted watermark.

Last, LSCs must be defined, to receive the watermark, and the embedding
method must be chosen in order to alter a few LSCs with the watermark: each
of this stages can be achieved in a lot of different ways.

Let us summarize, in a scheme, the above watermarking method.

\begin{center}
\includegraphics{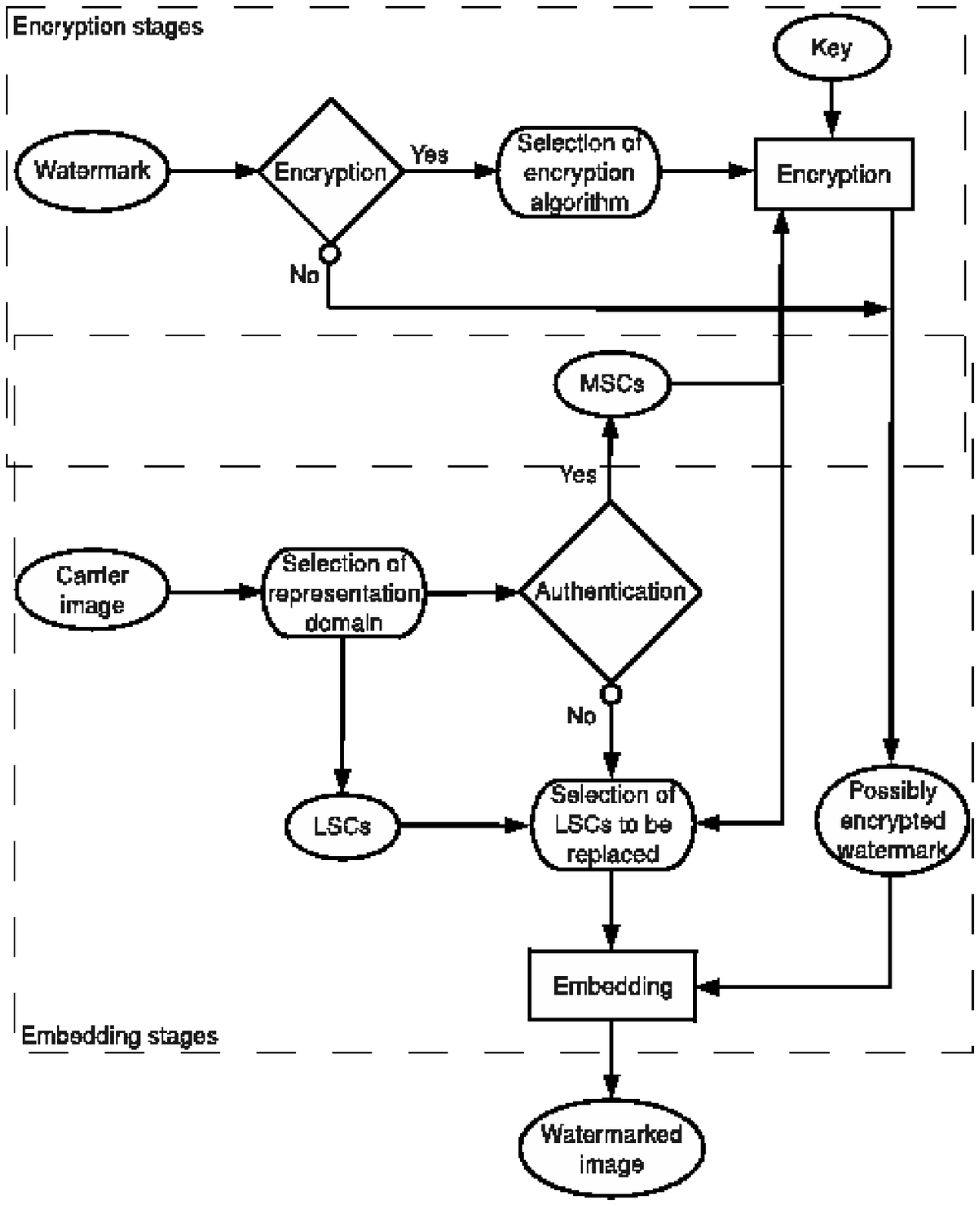}\\[0pt]
\emph{Fig 10.} The chaos-based watermarking algorithm
\end{center}

\section{Related works and contributions}

In~\cite{Wu2007} and \cite{Wu2007bis}, a new chaotic watermarking method is
presented, starting by the encryption $W_{e}$ of a watermark $W$, thanks to
the formula: $W_{e}=W\otimes X$, where $\otimes $ denotes the bitwise
exclusive or, and $X$ is the bitwise logistic map. Then, for each bit $b$ of
the encrypted watermark, a pixel of the carrier image is chosen to embed $b$%
, with 2-D Arnold cat map\footnote{%
2-D Arnold cat map is defined by 
\begin{equation}
\left\{ 
\begin{matrix}
X^{n+1} & = & X^{n}\,+\,Y^{n}\quad (\mathrm{mod}\ 1) \\ 
Y^{n+1} & = & X^{n}\,+\,2\,Y^{n}\quad (\mathrm{mod}\ 1)%
\end{matrix}%
\right. 
\end{equation}%
}, and the logistic map is used another time, translated as above into a
sequence of $\{4,5,6,7\}$, to choose which least significant bit (LSB) of
this pixel will be replaced by $b$. The watermark extraction is just the
inverse process of the embedding algorithm.

Other schemes for the choice of the LSBs have been proposed, the goal is
mostly to change the robustness against attacks and to make the detection of
the watermark more difficult. To carry out that hash function or digital
signature were used. The paper \cite{liu2007bis} is an example where the
weakness of the embedding process is needed. Other embedding domains were
also explored. DFT, DCT and DWT are three frequently used transformations.
For example, in \cite{Zhao2004}, logistic and Arnold cat maps are used with
neural networks to encrypt the watermark which is embedded in the wavelet
domain. Other examples of such embedding, using DWT and chaotic map, can be
found in \cite{Cong2006}, or in \cite{Dawei2004}.

\medskip

In the above mentioned papers, chaos is frequently employed in order to
ensure the strong authentication and secure encryption and uniform embedding
of the watermark. However, the question of how and what chaotic properties
are really necessary to achieve these goals is never elucidated. Moreover,
authentication and encryption are not always proposed.

In this study we try to ensure these objectives using a global and
comprehensive approach. It should be noticed that chaotic iterations, on
which our method is based, can be written in the field of discrete dynamic
systems:

\begin{equation*}
\left\{ 
\begin{array}{l}
x^{0}\in \mathcal{X} \\ 
x^{n+1}=f(x^{n})%
\end{array}%
\right. 
\end{equation*}%
where $(\mathcal{X},d)$ is a metric space (for a distance to be defined),
and $f$ is a continuous function (see \cite{Bahi2008}). Thus, it becomes
possible to study the topological behavior of those chaotic iterations.

Precisely, we have proved in \cite{Bahi2008} that, if the iterate function $f
$ is suitably chosen, then chaotic iterations generate a chaos in the
meaning of Devaney (its definition can be found in the Appendix, or in~\cite%
{Dev89}). Therefore, chaotic iterations as Devaney's topological chaos
satisfy sensitive dependence to the initial condition, unpredictability,
indecomposability and uniform repartition. Sensitivity to initial conditions
is used in authentication: in this case, watermark encryption and LSCs are
highly dependent to any changes of the carrier image. Unpredictability make
it impossible to determine whose LSCs have been altered. Last, the watermark
cannot be removed, even by cropping carrier image: it is theoretically
possible to make the authentication of a watermarked image, even by studying
just a part of it (Devaney's chaos definition contain the so called
transitivity property: the watermark can be found in any part of the carrier
image).

\section{A case study}

\label{CaseStudy} In this section, an application example of the above
chaotic watermarking method is given, and its robustness through attacks is
studied.

\subsection{Stages and details}

\subsubsection{Images description}

Carrier image is the so famous Lena, which is a 256 grayscale image, and the
watermark is the following $64\times 64$ pixels binary image:

\begin{center}
\begin{tabular}{cc}
\includegraphics[scale=0.5]{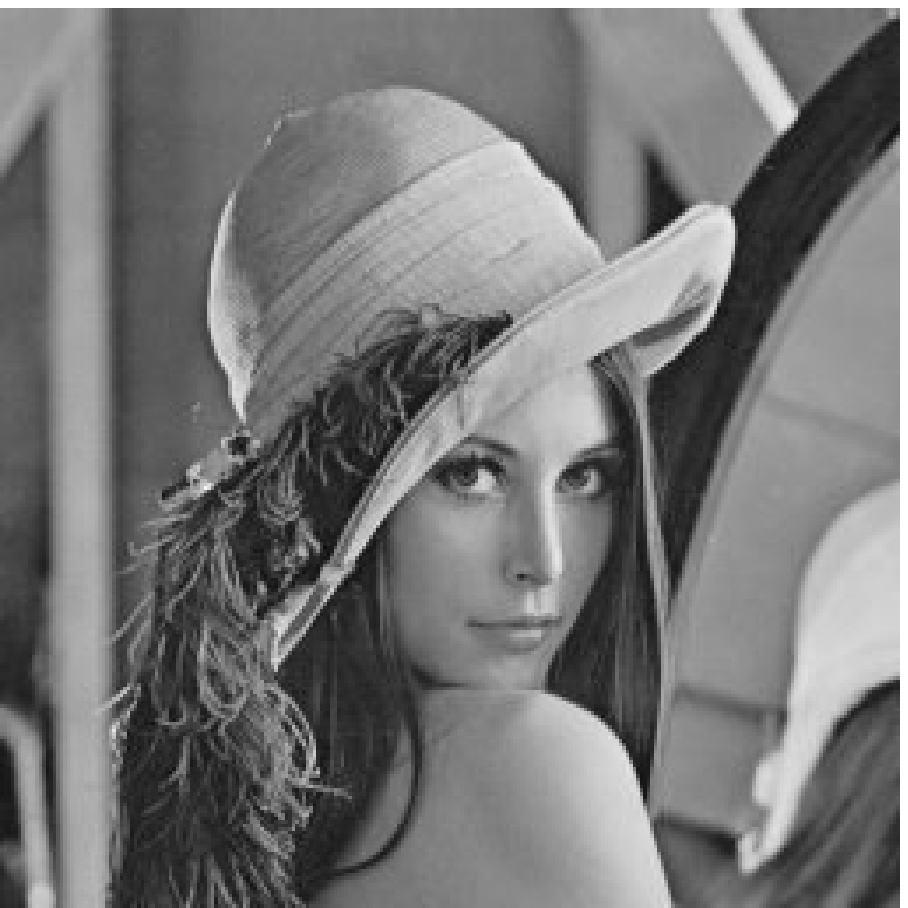} & \includegraphics{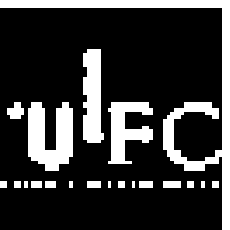} \\ 
\emph{Fig 2.} Lena (\emph{scale} 0.5) & \emph{Fig 3.} Watermark%
\end{tabular}
\end{center}

The embedding domain will be the spatial domain: selected MSCs are the four
most significant bits of each pixel, and LSCs are the three following bits
(a given pixel will at most be modified of four levels of gray, see \emph{%
Fig. 1a, 1b}).

\subsubsection{Encryption of the watermark}

\label{watermark encryption}

Let us explain how to encrypt the watermark by using chaotic iterations
defined by definition \ref{Def:Chaotic iterations} in section \ref{chaotic
iterations}. The initial state $x^{0}$ of the system is constituted by the
watermark, considered as a boolean vector. The iteration function is the
vectorial logical negation\footnote{%
The \emph{vectorial logical negation is defined by }%
\begin{equation*}
\begin{array}{rccc}
f_{0}: & \mathds{B}^{\mathsf{N}} & \longrightarrow & \mathds{B}^{\mathsf{N}}
\\ 
& (x_{1},\hdots,x_{\mathsf{N}}) & \longmapsto & (\overline{x_{1}},\hdots,%
\overline{x_{\mathsf{N}}}) \\ 
&  &  & 
\end{array}
\label{f0}
\end{equation*}%
} $f_{0}$, and the chaotic strategy $(S^{k})_{k\in \mathds{N}}$ will depends
on whether an authenticated watermarking method is desired or not, as
follows.

A chaotic boolean vector is generated by a number $T$ of iterations of a
logistic map ($(\mu ,U_{0})$ parameters will constitute the private key).
Then, in case of unauthenticated watermarking, the bits of the chaotic
boolean vector are grouped six by six, to obtain a sequence of integers
lower than 64, which will constitute the chaotic strategy. In case of
authentication, the bitwise exclusive or (XOR) is made between the chaotic
boolean vector and the MSCs, and the result is converted into a chaotic
strategy by joining its bits as above. Thus, the encrypted watermark is the
last boolean vector generated by the chaotic iterations. \newline
Let us give some examples of such an encryption ($T=5000$ iterations, $\mu
=4,X_{0}=0.61$):

\begin{center}
\begin{tabular}{ccc}
\includegraphics{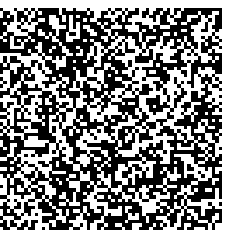} & \includegraphics{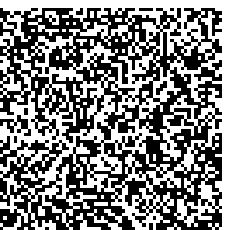} & %
\includegraphics{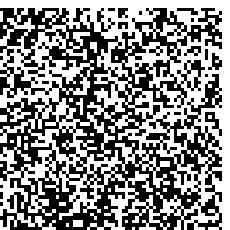} \\ 
\emph{Fig 4a.} Without authentication & \emph{Fig 4b.} With authentication & 
\emph{Fig 4c.} Differences \\ 
&  & between $4a$ and $4b$: 49,27 \% \\ 
&  & 
\end{tabular}
\end{center}

\subsubsection{Embedding of the watermark}

\label{Watermark embedding}

To embed the watermark, the sequence $(U^{k})_{k\in \mathds{N}}$ of altered
bits taken from the LSCs must be defined. To do so, the strategy $%
(S^{k})_{k\in \mathds{N}}$ defined in the encryption stage is used as
follows 
\begin{equation}
\left\{ 
\begin{array}{lll}
U^{0} & = & S^{0} \\ 
U^{n+1} & = & S^{n+1}+2\times U^{n}+n%
\end{array}%
\right.
\end{equation}

\begin{center}
\begin{tabular}{cc}
\includegraphics[scale=0.5]{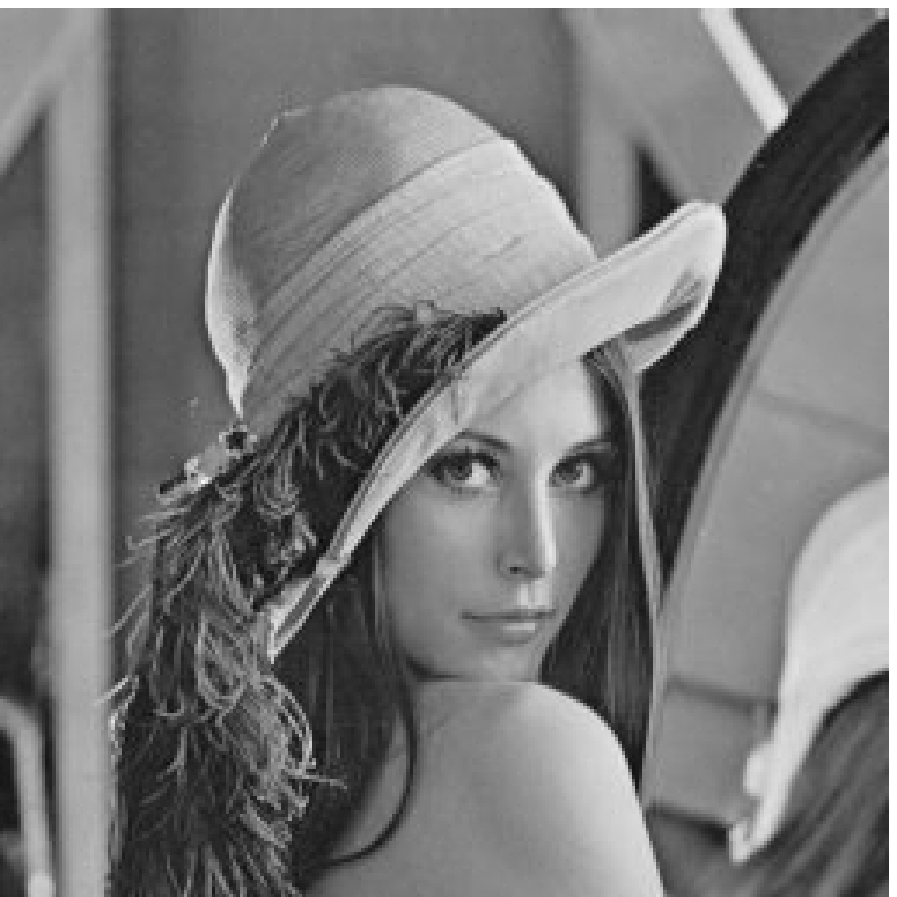} & %
\includegraphics[scale=0.5]{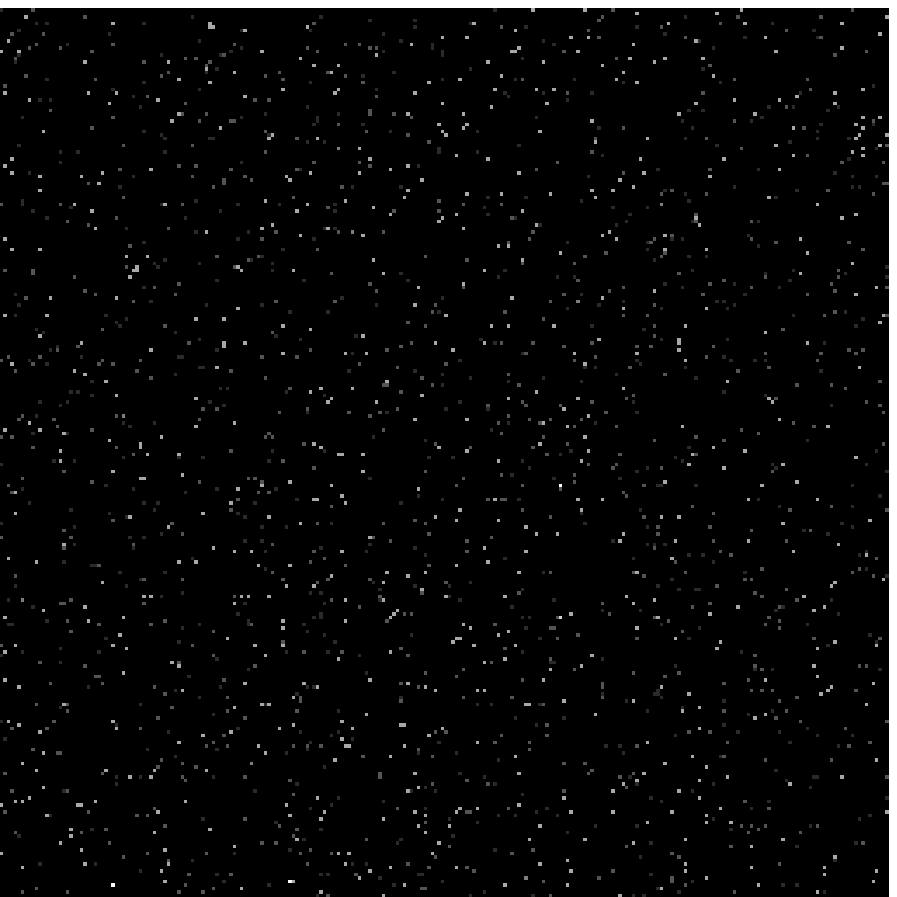} \\ 
\emph{Fig 5a.} Watermarked Lena & \emph{Fig 5b.} Differences between original
\\ 
(encryption, no authentication) & and watermarked Lena \\ 
PSNR = 54.78 & (pixel values increased by a factor 42) \\ 
& 
\end{tabular}
\end{center}

Remark that the map $\theta \mapsto 2\theta $ of the torus, which is a
famous example of topological Devaney's chaos \cite{Dev89}, has been chosen
to make $(U^{k})_{k\in \mathds{N}}$ high sensitive to the chaotic strategy.
As a consequence, $(U^{k})_{k\in \mathds{N}}$ is high sensitive to the
alteration of the MSCs: in case of authentication, any significant
modification of the watermarked image will conduct to a completely different
extracted watermark.

\subsection{Simulation results}

In what follows, as in subsection \ref{CaseStudy}\ref{watermark encryption}
the embedding domain is the spatial domain, logistic map with parameters $%
(\mu =4,$ $X_{0}=0.61)$ has been used to encrypt the watermark, MSCs are the
four firsts bits of each pixel (useful only in case of authentication), and
LSCs are the three next bits.

To prove the efficiency and the robustness of the proposed algorithm, some
attacks are applied to our chaotic watermarked image. For each attack, a
similarity percentage with the watermark is computed: this percentage is the
number of equal bits between the original and the extracted watermark, shown
as a percentage. Let us notice that a result less than or equal to $50\%$
implies that the image has probably not been watermarked.

\subsubsection{Geometric attacks}

\paragraph{Cropping attack.}

In this kind of attack, a watermarked image is cropped, such as:

\begin{center}
\includegraphics[scale=0.5]{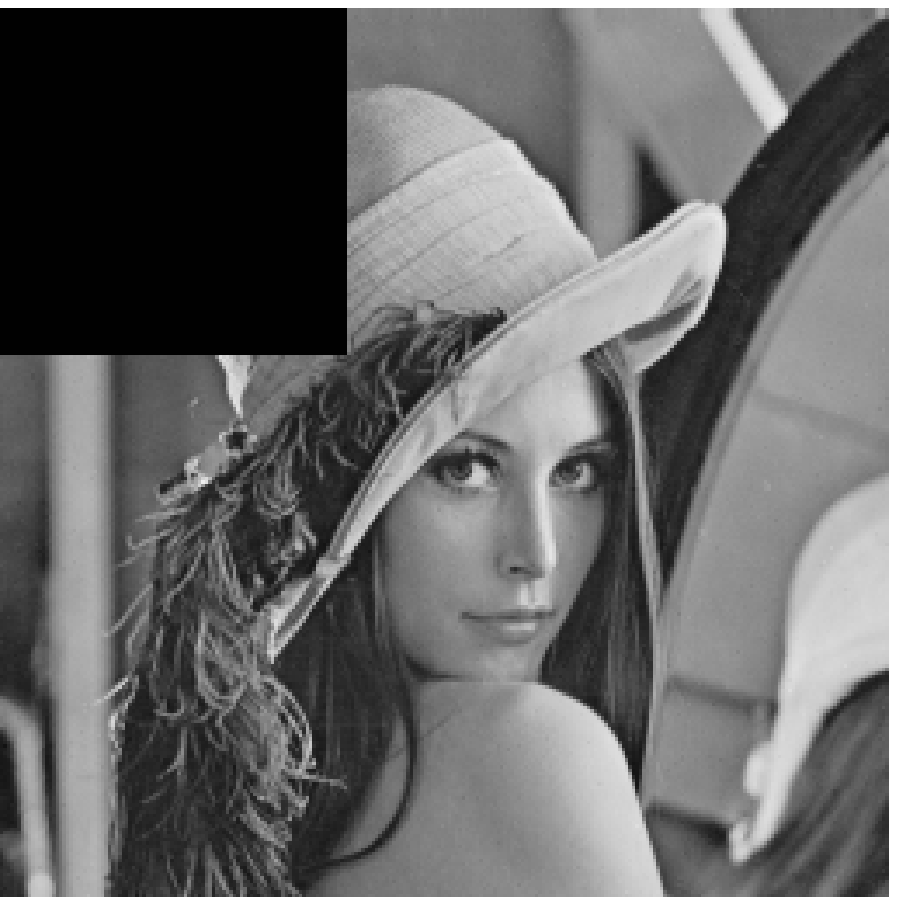}\\[0pt]
\emph{Fig 6.} Crop by $100 \times 100$ pixels.
\end{center}

In this case, the following results have been obtained...

\begin{center}
\begin{tabular}{|c||c|}
\hline
UNAUTHENTICATION & AUTHENTICATION \\ \hline\hline
\begin{tabular}{|c|c|}
\hline
Size (pixels) & Similarity percentage \\ \hline
10 & 99.08\% \\ 
50 & 97.31\% \\ 
100 & 92.43\% \\ 
200 & 70.75\% \\ \hline
\end{tabular}
& 
\begin{tabular}{|c|c|}
\hline
Size (pixels) & Similarity percentage \\ \hline
10 & 89.81\% \\ 
50 & 54.54\% \\ 
100 & 52.24\% \\ 
200 & 51.87\% \\ \hline
\end{tabular}
\\ \hline
\end{tabular}
\end{center}

In what follows, the decrypted watermarks are shown after a crop of 50
pixels, and after a crop of 10 pixels, in the authentication case.

\begin{center}
\begin{tabularx}{\linewidth/2}{ccc}
\includegraphics{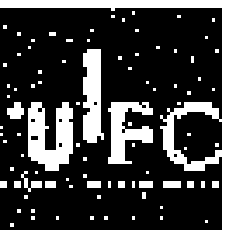} & \includegraphics{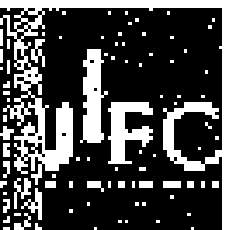}&
\includegraphics{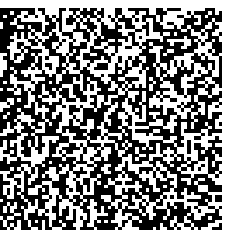} \\
 $50\times 50$ pixels  & $10 \times 10$ pixels & $50\times 50$ pixels \\
Unauthentication & \multicolumn{2}{c}{Authentication} \\
\end{tabularx}\\[0pt]
\emph{Fig 7.} Extracted watermark after a cropping attack.
\end{center}

In case of unauthentication, the watermark still remain after a cropping
attack: the desired robustness is reached. It can be noticed that cropping
sizes and percentages are rather proportional.

In case of authentication, even a small change of the carrier image (a crop
by $10 \times 10$ pixels) conduct to a really different extracted watermark.
In this case, any attempt to alter the carrier image will be signaled, image
is well authenticated.

\paragraph{Rotation attack.}

Let $r_\theta$ be the rotation around the center $128 \times 128$ of the
carrier image, by angle $\theta$. So, the transformation $r_{-\theta}\circ
r_\theta$ is applied to the watermarked image, which is altered as follows

\begin{center}
\includegraphics[scale=0.5]{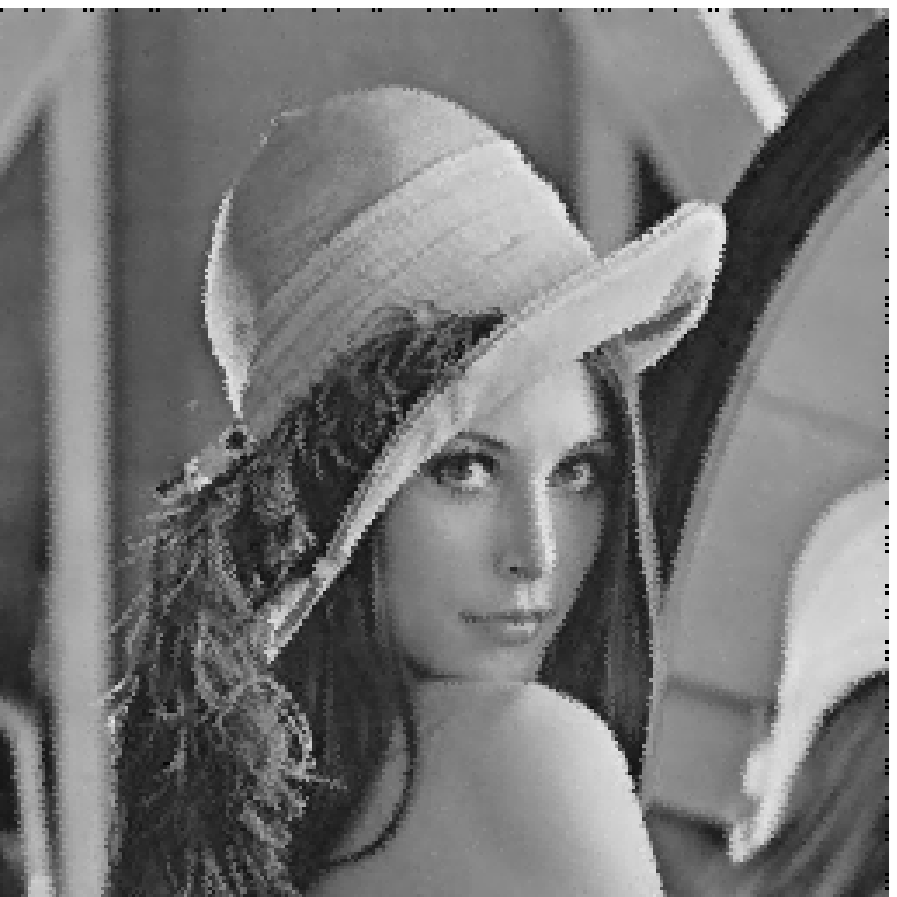}\\[0pt]
\emph{Fig 8.} After $r_{-25^{\circ}}\circ r_{25^{\circ}}$
\end{center}

In this case, the following results are obtained.

\begin{center}
\begin{tabular}{|c||c|}
\hline
UNAUTHENTICATION & AUTHENTICATION \\ \hline\hline
\begin{tabular}{|c|c|}
\hline
Angle (degree) & Similarity percentage \\ \hline
2 & 97.51\% \\ 
5 & 94.67\% \\ 
10 & 91.30\% \\ 
25 & 80.85\% \\ \hline
\end{tabular}
& 
\begin{tabular}{|c|c|}
\hline
Angle (degree) & Similarity percentage \\ \hline
2 & 70.01\% \\ 
5 & 59.47\% \\ 
10 & 54.51\% \\ 
25 & 50,21\% \\ \hline
\end{tabular}
\\ \hline
\end{tabular}
\end{center}

Decryption of the extracted watermark, after a rotation by $5^{\circ}$:

\begin{center}
\begin{tabular}{cc}
\includegraphics{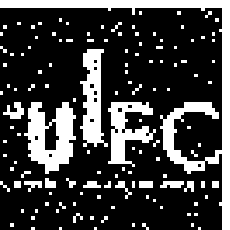} & %
\includegraphics{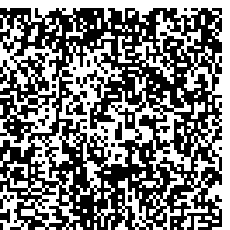} \\ 
\emph{Fig 9a.} Unauthentication & \emph{Fig 9b.} Authentication%
\end{tabular}
\end{center}

The same conclusion as above can be declaimed: this watermarking method
satisfies the desired properties.

\subsubsection{Other attacks}

\paragraph{JPEG compression.}

A JPEG compression is applied to the watermarked image, depending on a
compression level. Let us notice that this attack conducts to a change of
the working domain (from spatial to DCT domain). In this case, the following
results have been obtained:

\begin{center}
\begin{tabular}{|c||c|}
\hline
UNAUTHENTICATION & AUTHENTICATION \\ \hline\hline
\begin{tabular}{|c|c|}
\hline
Compression level & Similarity percentage \\ \hline
2 & 82.95\% \\ 
5 & 65.23\% \\ 
10 & 60.22\% \\ 
20 & 53.17\% \\ \hline
\end{tabular}
& 
\begin{tabular}{|c|c|}
\hline
Compression level & Similarity percentage \\ \hline
2 & 54.39\% \\ 
5 & 53.46\% \\ 
10 & 50.14\% \\ 
25 & 48.80\% \\ \hline
\end{tabular}
\\ \hline
\end{tabular}
\end{center}

A very good authentication through JPEG attack is obtained. As for the
unauthentication case, the watermark still remain after a compression level
equal to 10: this is a good result for a spatial embedding.

\paragraph{Gaussian noise.}

Watermarked image can be attacked too by the addition of a Gaussian noise,
depending on a standard deviation. In this case, the following results have
been obtained:

\begin{center}
\begin{tabular}{|c||c|}
\hline
UNAUTHENTICATION & AUTHENTICATION \\ \hline\hline
\begin{tabular}{|c|c|}
\hline
Standard deviation & Similarity percentage \\ \hline
1 & 74.26\% \\ 
2 & 63.33\% \\ 
3 & 57.44\% \\ 
5 & 51.26\% \\ \hline
\end{tabular}
& 
\begin{tabular}{|c|c|}
\hline
Standard deviation & Similarity percentage \\ \hline
1 & 52.05\% \\ 
2 & 50.95\% \\ 
3 & 49.65\% \\ 
5 & 49.43\% \\ \hline
\end{tabular}
\\ \hline
\end{tabular}
\end{center}

We obtain the same conclusion as above.

\section{Discussion and future work}

In this paper, a new way to generate watermarking methods is proposed. The
new procedure depends on a general description of the carrier medium to
watermark in terms of the significance of some coefficients we are called
MSC and LSC in definition 3. We proposed a general and comprehensive
algorithm which recalls the different choices and possibilities that our
approach offers.

\medskip

This encryption and the selection of coefficients to alter are based on
chaotic iterations \cite{Bahi2008} which generate a topological chaos in the
sense of Devaney \cite{Dev89}. So, the properties of topological chaos are
satisfied by our algorithm, such as sensitivity to initial conditions,
uniform repartition (thanks to the transitivity) and unpredictability. Thus,
the proposed algorithm possesses the desirable properties expected in good
watermarking algorithms. For example, a strong authentication of the carrier
image is satisfied thanks to the sensitivity, the unpredictability property
implies security, and the resistance against attacks and discretion of the
watermark are consequences of the transitivity property.

\medskip

The algorithm has been evaluated through attacks, and all of the expected
results have been experimentally obtained. Choices that have been made in
this first study are simple: spatial domain, negation function for the
iteration function. The aim was not to find the best watermarking method
generated by our general algorithm, but to give a simple illustration
example, to prove its feasibility.

\medskip

In future work, in order to increase authentication and resistance to
attacks, other choices of the iteration function and the chaotic strategy
will be studied and compared. In addition, frequency domain representations
will be used to select the MSCs and LSCs. Moreover others properties ensured
by chaotic iterations, such as entropy, will be shown and their role in
watermarking algorithms will be explained..

\bibliographystyle{habbrv}
\bibliography{/home/guyeux/Documents/These/mabase.bib}

\bigskip

\section*{Appendix: Devaney's chaotic dynamic systems}

Consider a metric space $(\mathcal{X},d)$, and a continuous function $f:%
\mathcal{X}\longrightarrow \mathcal{X}$.

\begin{definition}
$f$ is said to be \emph{topologically transitive} if, for any pair of open
sets $U,V \subset \mathcal{X}$, there exists $k>0$ such that $f^k(U) \cap V
\neq \varnothing$.
\end{definition}

\begin{definition}
An element (a point) $x$ is a periodic element (point) for $f$ of period $%
n\in \mathds{N},$ if $f^{n}(x)=x$. The set of periodic points of $f$ is
denoted $Per(f).$
\end{definition}

\begin{definition}
$(\mathcal{X},f)$ is said to be \emph{regular} if the set of periodic points
is dense in $\mathcal{X}$, 
\begin{equation*}
\forall x\in \mathcal{X},\forall \varepsilon >0,\exists p\in
Per(f),d(x,p)\leqslant \varepsilon .
\end{equation*}
\end{definition}

\begin{definition}
\label{sensitivity} $f$ has \emph{sensitive dependence on initial conditions}
if there exists $\delta >0$ such that, for any $x\in \mathcal{X}$ and any
neighborhood $V$ of $x$, there exists $y\in V$ and $n\geqslant 0$ such that $%
|f^{n}(x)-f^{n}(y)|>\delta $.

$\delta$ is called the \emph{constant of sensitivity} of $f$.
\end{definition}

Let us recall the definition of a chaotic topological system, in sense of
Devaney~\cite{Dev89}:

\begin{definition}
$f:\mathcal{X}\longrightarrow \mathcal{X}$ is said to be \emph{chaotic} on $%
\mathcal{X}$ if, and only if
\begin{enumerate}
\item $f$ has sensitive dependence on initial conditions,
\item $f$ is topologically transitive,
\item $(\mathcal{X},f)$ is regular.
\end{enumerate}
\end{definition}

\end{document}